\documentclass[aps,prb,amsmath,amssymb,twocolumn,showpacs,floatfix]{revtex4}

\usepackage{graphicx}
\usepackage{epsfig}
\usepackage{amsmath}
\usepackage{multirow}

\begin{document}

\title{Quantum Monte Carlo study of a vortex in superfluid $^4$He and search for a vortex state in the solid}

\author{D.E. Galli$^1$, L. Reatto$^2$, and M. Rossi$^3$}
 \affiliation{
 $^1$ Dipartimento di Fisica, Universit\`a degli Studi di Milano, via Celoria 16, 20133 Milano, Italy\\
 $^2$ via Bazzini 20, 20133 Milano, Italy\\
 $^3$ Dipartimento di Fisica e Astronomia ``Galileo Galilei'', Universit\`a degli Studi di Padova, via Marzolo 8,  35131 Padova, Italy}

\date{\today}

\begin{abstract}
We have performed a microscopic study of a straight quantized vortex line in 
three dimensions in condensed $^4$He at zero temperature using the Shadow 
Path Integral Ground State method and the fixed-phase approximation.
We have characterized the energy and the local density profile around the 
vortex axis in superfluid $^4$He at several densities, ranging from below the 
equilibrium density up to the overpressurized regime.
For the Onsager-Feynman (OF) phase our results are exact and represent a 
benchmark for other theories.
The inclusion of backflow correlations in the phase improves the description 
of the vortex with respect to the OF phase by a large reduction of the 
core energy of the topological excitation.
At all densities the phase with backflow induces a partial filling of the 
vortex core and this filling slightly increases with density.
The core size slightly decreases for increasing density and the density 
profile has well defined density dependent oscillations whose wave vector is 
closer to the wave vector of the main peak in the static density response 
function rather than to the roton wave vector.
Our results can be applied to vortex rings of large radius $R$ and we find
good agreement with the experimental value of the energy as function of $R$ 
without any free parameter.
We have studied also $^4$He above the melting density in the solid phase
using the same functional form for the phase as in the liquid.
We found that off-diagonal properties of the solid are not qualitatively 
affected by the velocity field induced by the vortex phase, both with and 
without backflow correlations.
Therefore we find evidence that a perfect $^4$He crystal is not a marginally
stable quantum solid in which rotation would be able to induce off--diagonal 
long--range coherence. 
\end{abstract}


\maketitle

\section{INTRODUCTION}
\label{s:intro}

Topological excitations represent a class of excitations of fundamental 
interest in many ordered phases in condensed matter like Bose/BCS--condensed 
quantum fluids, superconductors, crystals or nematic liquid crystals. 
Starting from the works by Onsager\cite{onsa} and by Feynman\cite{feyn} a
widely studied example of a topological excitation is a vortex line in 
a Bose superfluid, in particular in superfluid $^4$He. 
Vortices play a fundamental role in many superfluid phenomena, for instance 
the behavior of a superfluid under rotation or the value of the critical velocity
for the onset of dissipation in many cases are determined by vortex nucleation. 
Addressing specifically superfluid $^4$He almost all the studies of vortices
are based on simplified models in which vorticity turns out to be localized 
along mathematical lines, more precisely the phase of the wave function (wf) 
is assumed to be additive in the phase of each particle, the so called 
Onsager-Feynman (OF) form.
Within this approximation the vorticity field has a singularity along a line,
the vortex core, where the density vanishes and the velocity diverges.
This behavior is found, for instance, with the Gross-Pitaevskii (GP) equation 
\cite{pita,gros} or with the Biot-Savart model of vortex filaments.\cite{saff} 
Such models can be a reasonable approximation for weakly interacting particles
like cold bosonic atoms. 
For a strongly correlated system like superfluid $^4$He that approximation is 
questionable because single particle phase additivity is incompatible with the
presence of interparticle correlations that lead to backflow effects. 
Still, also in superfluid $^4$He, most of the studies are based on models 
with singular vorticity.
A justification for this is that the healing length $\xi$ of the superfluid 
order parameter is of order of one {\AA}ngstrom, orders of magnitude smaller 
than the typical inter-vortex distance.
Therefore in most instances the flow field of a given vortex system is equal 
to that given by classical incompressible hydrodynamics with the single 
constraint that the circulation $\kappa$ around each vortex filament is 
quantized in unit of Plank's constant over particle mass, $\kappa=h/m$. 
This explains why only few studies have addressed the local structure of a 
vortex in superfluid $^4$He beyond the singular vorticity models.

The previous perspective is changing due to the intense experimental and 
theoretical interest in vorticity phenomena at low temperature 
\cite{fett,tsub} where the normal component of the superfluid essentially 
vanishes. 
Under such conditions diffusion and decay of a vortex tangle, as observed
experimentally,\cite{lanc} must be dominated by reconnection of vortices, the 
only mechanism that can change the topology of the vortex system in absence of 
dissipation. 
Computations\cite{reco} based on the GP equation show that 
reconnections take place when the distance between two vortex cores is of 
order of the healing length $\xi$. 
On the basis of the GP equation the local density vanishes at the vortex line
and the density reaches smoothly the bulk value within a distance of order of 
$\xi$, whereas it is known that interparticle correlations lead to density 
oscillations as function of distance from the vortex axis.\cite{ches} 
It should be noted that when the GP equation is used to study the elementary 
excitations of the system, the bulk excitations consist of phonons joining 
monotonically free particle behavior at large wave vectors and that roton 
excitations are not present.
Rotons are excitations arising in presence of strong interparticle 
correlations.\cite{feyC,cecc,rota,maur} 
The nature of the bulk excitations can be relevant in connection to vortex 
reconnections because there is evidence that a reconnection event is 
associated with emission of bulk excitations, in addition to vortex 
oscillations (Kelvin waves).\cite{kelv}                                           
More precisely studies based on the GP equation\cite{lead,zucc} have shown that 
vortex reconnection events generate rarefaction waves, i.e. bulk sound waves. 
This suggests that a realistic study of reconnections in superfluid $^4$He 
needs to start from a good model of the vortex core and, at the same time, of
the excitations of bulk superfluid $^4$He with a proper treatment not only of
phonons but also of rotons.\cite{ogaw} 
The more so because on the basis of density functional theory it has been 
shown\cite{dalf,vill} that the oscillation of the density profile around the vortex 
core seems to be related to the roton excitations.
Recent progress\cite{bew1,bew2,zmee} in the visualization at a local level of quantum      
vorticity should allow studies of vortex reconnections and quantum turbulence at
a level of detail not available before so that advances in theoretical modeling
are called for.

In the literature only very few studies are present of the core of a vortex 
in superfluid $^4$He based on microscopic theory that goes beyond the mean field
singular vortex behavior. 
In three dimensional (3D) $^4$He the only study is the one\cite{vit0,vit2} based
on variational theory with shadow wave function (SWF). 
Another study was presented of a vortex in superfluid $^4$He in mathematical two
dimensions (2D) based on the so called fixed phase quantum Monte Carlo (FP-QMC).\cite{orti}
Also FP-QMC is a variational approach but it goes beyond the approach of 
Ref.~\onlinecite{vit0,vit2} because, for an assumed form of the phase of the wf,
the real part of the wf is computed exactly. 
In these works \cite{vit0,vit2,orti} the global vortex phase is not additive in 
the single particle phases but it contains also pair or more complex contributions. 
Commonly one says that backflow effects are taken into account. 
This term has its origin in the Feynman-Cohen theory \cite{feyC} of rotons in which 
the phase of such momentum carrying excited state has not only single particle 
contributions, like in the Feynman theory,\cite{fey2} but also contributions 
depending on the relative positions of pairs of neighboring atoms. 
Such pair contributions are needed in order to guarantee local conservation of 
matter and there is some similarity with the backflow effects in classical 
hydrodynamics. 
A visualization of such roton backflow can be found in Ref.~\onlinecite{cecc} where 
roton wave packets have been studied by a microscopic theory. 
As far as we know, no study of a vortex in the 3D $^4$He based on
advanced QMC methods has been performed yet and this is the problem that we 
address in the present study of a straight vortex line in condensed $^4$He via the
shadow path integral ground state\cite{spigs1,spigs2} (SPIGS) method with fixed
phase approximation.
We have studied a vortex line in superfluid $^4$He over an extended pressure 
range, from a density below equilibrium close to the spinodal up to a 
density in the deeply overpressurized liquid $^4$He at a density about 15\% 
above the freezing density.

Following our recent work on 2D {\it solid} $^4$He,\cite{2Dvor} here we have 
studied also a possible vortex state in 3D solid $^4$He by considering 
different phases with FP-QMC. 
This investigation is motivated by the presence of
phenomenological models based on vortices\cite{ande} to explain the possible
superfluid response of solid $^4$He,\cite{newrev} i.e. supersolidity.
In fact, the vortex model has been used to interpret several 
experiments.\cite{kubo1,kubo2,kim1,kim2}
Supersolidity in solid $^4$He is a debated question\cite{newrev} and there is no
shared consensus on the supersolid nature of $^4$He.
A point that is firmly established by QMC computations is that in an ideal 
perfect (i.e. with no defects, the so called commensurate crystal) $^4$He 
crystal the condensate fraction and the superfluid density are zero at 
finite\cite{prok,cipo} and even at zero temperature.\cite{fili,miarev}
In absence of phase coherence our choices for the phase of the wave function
(Eq.~\eqref{OFphase} and \eqref{BFphase}) have little justification.
It has been put forward the conjecture that the non-supersolid state of $^4$He 
could be marginally stable (i.e. almost any deviation from the perfect crystal 
would lead to a superfluid response).\cite{newrev,miarev} 
Then, the present computation is relevant to infer if the centrifugal barrier
associated to the flow field of a vortex line is able to induce in an ideal
perfect crystal off-diagonal long--range order.
  
This paper is organized as follows:
in section \ref{s:meth} we discuss the fixed phase approximation applied to the 
simulation of a vortex line with the path integral ground state method; 
our results for a straight vortex line in the liquid phase of $^4$He are shown in section
\ref{s:res} and how our results can be applied to a vortex ring of large diameter is 
discussed in section \ref{s:ring}. 
In section \ref{s:solid} we present the results for the solid phase and 
section \ref{s:concl} contains our conclusions.

\section{Methods}
\label{s:meth}

Dealing with vortices in a Bose fluid is an unresolved problem for exact 
microscopic ab-initio methods, and it calls for some approximations or 
assumptions.
For instance,  in order to describe a straight vortex line, the many--body 
wf $\Psi(R)$ has to be an eigenstate of the angular momentum operator 
$\hat{\mathcal{L}}_z$ with eigenvalues $\hbar Nl$, being $l=1,2,\dots$ the 
quanta of circulation; 
this requires the presence of a phase and thus one deals with a complex wave 
function.
It seems really tempting, following the well established route for the ground 
state, to try improving a variational ansatz by exploiting the projection 
ability of exact Quantum Monte Carlo (QMC) techniques.
Unfortunately the complex nature of the wf rules out an exact
implementation of QMC methods due to the presence of a phase problem. 
The most followed recipe is then to overcome the sign problem by releasing on 
the exactness of QMC techniques and improving only some aspects of the trial 
wf, keeping others aspects at a variational level.
This is the case of approximations like fixed phase\cite{orti,sola} or 
fixed node.\cite{gior}
Such approximations are viable also with finite temperature methods that do not
involve explicitly the wf, like path integral Monte 
Carlo.\cite{drae}

In full generality, the many body wf can be written as
$\Psi(R)=e^{i\Omega(R)}\Psi_0(R)$, where $\Omega(R)$ is the phase and 
$\Psi_0(R)$ is the modulus of the wf, and 
$R=\{\vec r_1, \vec r_2,\dots,\vec r_N \}$ represents the coordinates of the $N$
particles composing the system.
$\Psi(R)$ describes a stationary quantum state if it is a solution of the time 
independent Schr\"odinger equation $\hat{\mathcal{H}}\Psi(R) = E\Psi(R)$,
from which two coupled differential equations for $\Omega(R)$ and $\Psi_0(R)$ 
are readily obtained.
The fixed phase approximation\cite{orti} consists in assuming a given 
functional form for the phase $\Omega(R)$ and in solving the remaining 
differential equation for the modulus $\Psi_0(R)$:
\begin{equation}
 \label{schremod}
  \left[-\frac{\hbar^2}{2m}\sum_{i=1}^N\nabla_i^2 +
        V_\Omega(R) + V(R)\right]\Psi_0(R) = E_\Omega\Psi_0(R)
\end{equation}
where $V(R)$ represents the interatomic potential of the system and
$V_\Omega(R)$ reads
\begin{equation}
 V_\Omega(R) = \frac{\hbar^2}{2m}\sum_{i=1}^N 
                \left(\vec\nabla_i\Omega(R)\right)^2\ .
\end{equation}
Solving Eq.~\eqref{schremod} is equivalent to solve the original time 
independent Schr\"odinger equation for the $N$ particles with the extra 
potential term $V_\Omega(R)$.
Equation \eqref{schremod} can now be solved with one of the QMC methods that give
the exact energy and other properties of the system.
It can be proved that the fixed phase method provides a variational upper bound
for the lowest energy state among the wave functions having the assumed phase 
$\Omega(R)$.\cite{orti}

In the case of a straight vortex line, the simplest possible choice for the phase 
is the well known Onsager-Feynman (OF) phase:\cite{feyn}
\begin{equation}
 \label{OFphase}
 \Omega^{\rm OF}(R) = \sum_{i=1}^N\theta_i
\end{equation}
where $\theta_i$ is the azimuthal coordinate of the $i^{\rm th}$ particle with
respect to the vortex axis.
$\Omega^{\rm OF}(R)$ gives rise to an irrotational flow field everywhere but on
the vortex axis where the velocity field diverges and the quantized vorticity is
localized on this axis.
The extra potential term $V_\Omega(R)$ in Eq.~\eqref{schremod} is given by the
standard centrifugal barrier:
\begin{equation}
 \label{OFpot}
 V_\Omega^{\rm OF}(R) = \sum_{i=1}^Nv_\Omega^{\rm OF}(r_i)
\end{equation}
with
\begin{equation}
 v_\Omega^{\rm OF}(r_i) = \frac{\hbar^2}{2m}\frac{1}{r^2_i}
\end{equation}
where $r_i$ is the radial cylindrical coordinate of the $i^{\rm th}$ particle.
Due to the divergence of $V_\Omega^{\rm OF}(R)$ when a particle approaches the 
vortex axis, the local density has to vanish on the vortex line in order to have
finite kinetic energy.
The OF recipe, which provides a vortex line with an hollow core and localized
vorticity, has been largely employed to predict the properties of a vortex line
in  the ground state of bulk $^4$He via integral equation methods,\cite{ches} 
density functional\cite{dalf} and variational Monte Carlo (VMC),\cite{vit2,sad1},
and some of these techniques have been applied with good results also to 
Bose-condensed gases.\cite{nils} 

A way to improve the OF ansatz is taking into account the so called back-flow 
(BF) correlations\cite{orti} such that $V_\Omega(R)$ is no more a sum of single
particle terms and, by a proper choice of the phase, the velocity field can remain
finite everywhere, also at the vortex core, and the vorticity is no more
localized.
With the choice of Ref.~\onlinecite{orti} for the phase, it is found that BF
correlations lower the vortex energy compared to the OF choice (that computation
covers only 2D $^4$He \cite{orti,gior}) and, more important, it has a dramatic 
effect on another property of the system: the vortex core turns out to be no 
more empty\cite{orti,note} but the density is finite even inside the core.
Similar results in 3D have been also reached with shadow wave function (SWF) 
variational technique: the lowest energy state has a partially filled core with
distributed vorticity over a radius of about 1~\AA.\cite{vit2,sad1,sad2}

In the present computation we assume for $\Omega(R)$ of a straight vortex line a 
form that is an extension in 3D of the form studied by Ortiz and Ceperley in 
Ref.~\onlinecite{orti} in 2D.
Our BF phase $\Omega^{\rm BF}(R)$ reads 
\begin{equation}
 \label{BFphase}
 \Omega^{\rm BF}(R) = \sum_{j=1}^N 
                     ln\left(\frac{A_j+iB_j}{\sqrt{A_j^2+B_j^2}}\right) 
\end{equation} 
with $A_j=x_j + k\sum_{l\neq j}f(|\vec{r}_{jl}|,r_j,r_l)(x_j-x_l)$ and
$B_j=y_j + k\sum_{l\neq j}f(|\vec{r}_{jl}|,r_j,r_l)(y_j-y_l)$.
$f(|\vec{r}_{jl}|,r_j,r_l)=\exp{-(\alpha |\vec{r}_{jl}|^2 + \gamma (r_j^2 + r_l^2))}$
is the BF function\cite{orti} characterized by the variational parameters 
$k$, $\alpha$ and $\gamma$, and $|\vec{r}_{jl}|=|\vec{r}_j-\vec{r}_l|$.
In Eq.~\eqref{BFphase} we have used cartesian coordinates with the $z$ axis
taken along the vortex axis and $r_j=\sqrt{x_j^2+y_j^2}$.
The wf $\Psi(R)$ constructed with $\Omega^{\rm BF}(R)$ has some analogy with
the Feynman-Cohen wf for the phonon-roton excited states.\cite{feyC}
With this choice of the phase, the extra potential term in Eq.~\eqref{schremod}
reads
\begin{equation}
 \label{BFpot}
 V_\Omega^{\rm BF}(R) = \sum_{i=1}^Nv_{\Omega i}^{\rm BF}(R)
\end{equation}
with
\begin{eqnarray}
 v_{\Omega i}^{\rm BF}(R) &=& \frac{\hbar^2}{2m}\left(\frac{A_i\nabla_i B_i-B_i\nabla_i A_i}{A_i^2 + B_i^2}\right.\nonumber\\
                        &&  \left.+\sum_{l\neq i}\frac{A_l\nabla_i B_l-B_l\nabla_i A_l}{A_l^2 + B_l^2}\right)^2\ . 
\end{eqnarray}
For comparison purpose we have also performed computations for zero backflow 
(i.e. for $k=0$) so that one recovers the OF phase.\eqref{OFphase}

We face the task of solving \eqref{schremod} with the extra potential 
$V^{\rm OF}_\Omega(R)$ and $V^{\rm BF}_\Omega(R)$ with
the SPIGS method,\cite{spigs1,spigs2} which allows to obtain the exact lowest 
eigenstate of a given Hamiltonian and the exact correlation functions by 
projecting in imaginary time $\tau$ with the operator $e^{-\tau\hat H}$ a 
SWF\cite{vit1} taken as a trial wf.
As SWF we have used the optimized form\cite{mcfar} for bulk $^4$He.
It has been verified\cite{pata} that the SPIGS method is unbiased by the choice
of the trial wf, but a good choice of the trial wf is important in order to 
accelerate convergence as function of imaginary time and to reduce fluctuations.
Therefore in a SPIGS computation the only inputs are the interparticle potential
and the approximation for the imaginary time propagator.\cite{pata}
By a proper choice of the propagator the resulting errors can be reduced below 
the statistical uncertainty of the computation.
Notice that as trial wf we have used a wf for the bulk system which is uniform.
The non uniformity induced by the vortex phase is exclusively due to the action 
of the imaginary time evolution operator $e^{-\tau\hat H}$, so that by 
construction no bias is introduced in the computation once the phase factor has 
been chosen.
This is a great advantage of a PIGS computation compared, for instance, to the 
case of a computation with Green function and diffusion MC as in 
Refs.~\onlinecite{orti,gior}.  

As He--He interatomic potential we have considered the HFDHE2 Aziz potential\cite{aziz} 
and for the imaginary time propagator we have employed the primitive approximation.
The chosen time step is $\delta\tau = 1/640$ K$^{-1}$ and the total projection 
time is $\tau = 0.5$ K$^{-1}$, which represent a good compromise between 
accuracy and computational cost.

A technical difficulty with both \eqref{OFpot} and \eqref{BFpot} is represented
by their long range character which complicates the use of a finite simulation 
cell with periodic boundary conditions (pbc).
Often this problem has been overcome by simulating the system in a finite 
bucket,\cite{orti,vit2} but this has the drawback that two inhomogeneities are 
present at the same time, the one due to the vortex and the one due to the 
confining well.
Another possibility is to study a vortex-antivortex lattice in the bulk system
so that no large scale flow field is present and one can use pbc.\cite{sad1}
This approach brings in computational complications and it is difficult to 
implement with our form of BF.
On the other hand the vortex-antivortex lattice computation has shown that BF 
modifies the velocity flow field from the one of OF only at very short distance
from the core, below about 1~\AA.
Since we are interested in the characterization of local properties around the 
vortex core we have adopted a shortcut: we have simply smoothed the extra 
potentials $V_\Omega^{\rm OF}$ and $V_\Omega^{\rm BF}$ far from the core 
multiplying the terms $v_\Omega^{\rm OF}$ and $v_{\Omega i}^{\rm BF}$ in
\eqref{OFpot} and \eqref{BFpot} by the following function:
\begin{equation}
 \label{chifunc}
 \mathcal{S}(r)=\left\{
  \begin{array}{ll}
  1                             & r <\Delta            \\
  e^{-(r-\Delta)^2/(r-L/2)^2}   & \Delta\leq r\leq L/2 \\
  0                             & r > L/2              \\
  \end{array}
  \right.
\end{equation}
($L$ being the side of the simulation box) so that standard pbc can be applied.
With this choice, the extra potential is equivalent to \eqref{OFpot} and 
\eqref{BFpot} for $r<\Delta$, and gently smoothed to zero  by $\mathcal{S}(r)$ in the range 
$\Delta\leq r\leq L/2$.
The provided $\Psi(R)$ is no more an exact eigenstate of $\hat{\mathcal{L}}_z$ 
but we have verified that the local quantities we want to characterize, like the
integrated energy around the vortex line and the density around the core, are 
not affected by the value of $\Delta$, once it is taken sensibly greater than 
the core width.
The results shown in the next sections correspond to the choice $\Delta=8$ \AA.

\section{Vortex line in superfluid $^4$He}
\label{s:res}

We have studied a straight vortex line in bulk superfluid $^4$He at $T=0$ K at 
four different densities: near equilibrium density, $\rho=0.0218$ \AA$^{-3}$, near 
the freezing density, $\rho=0.026$ \AA$^{-3}$ ($P=25$ bar), below the equilibrium 
density, $\rho=0.02$ \AA$^{-3}$ ($P=-6$ bar), and finally, well above the freezing 
density, for a metastable overpressurized liquid, $\rho=0.03$ \AA$^{-3}$ ($P=71$ bar).
The reported values of pressure are derived from theoretical equation of state\cite{over}
for the adopted He-He interatomic potential.
The initial particle configurations for the metastable disordered computation have 
been obtained by rescaling to the desired density configurations from a simulation
performed below freezing;
in this way, as shown in Ref.~\onlinecite{over}, even at $\rho=0.03$ \AA$^{-3}$
the system remains disordered allowing for the characterization of the properties 
of the overpressurized phase.
The number of particles in the simulation box was taken to be $N=336$, which is large 
enough to ensure negligible size effects.
The adopted BF phase function, Eq.~\eqref{BFphase}, has three backflow parameters, 
the amplitude $k$, the interparticle range $\alpha$ and the particle-vortex range
$\gamma$ and these have been determined by minimization of the total energy of the
vortex.
Computations at $\rho=0.0218$ \AA$^{-3}$ have given $\alpha=0.1377$ \AA$^{-2}$ and
$\gamma=0.0765$ \AA$^{-2}$ (which correspond to a length scale of about $1.9$ and
$2.5$ \AA~ respectively ); these optimal values of $\alpha$ and $\gamma$ 
turn out to be equal to those obtained for the 2D system.\cite{orti}
Therefore we have retained these values also for all the other densities 
considered here.
Given this value for $\gamma$, we expect that the effect of the backflow will be 
restricted mainly within about 2.5 \AA~ from the vortex line.
The optimization of the backflow parameter $k$ gave the following optimal values:
$k=0.8$ at $\rho=0.02$ \AA$^{-3}$, $k=0.7$ at $\rho=0.0218$ \AA$^{-3}$, $k=0.6$ at
$\rho=0.026$ \AA$^{-3}$ and $k=0.6$ at $\rho=0.03$ \AA$^{-3}$.
When $k=0$ no vortex backflow effect is present and one recovers the OF phase
Eq.~\eqref{OFphase} which has no variational parameters.

\begin{figure*}
 \includegraphics*[width=15cm]{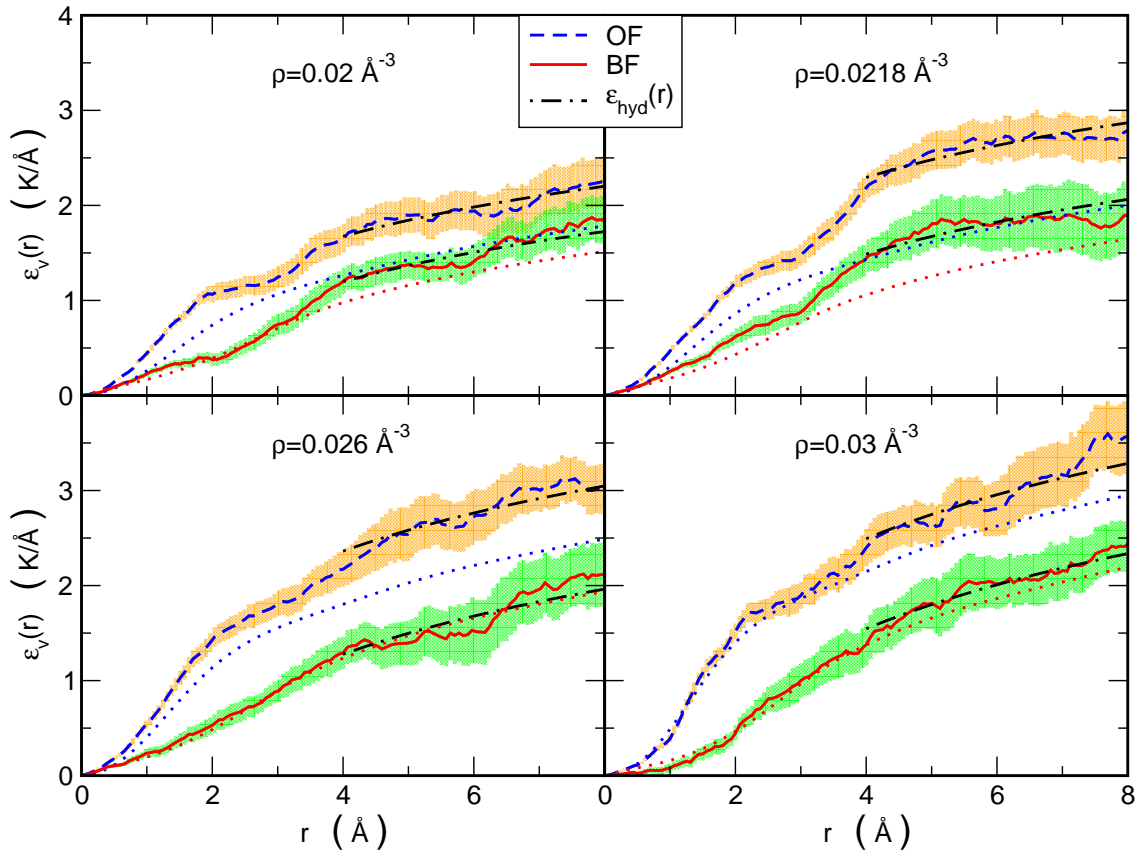}
 \caption{\label{fig1}
 (Color online) Integrated energy up to a distance $r$ of a straight vortex line per 
 unit length, $\varepsilon_v(r)$, at different densities in liquid phase as a function $r$
 Dashed line: Onsager-Feynman phase. 
 Solid line: Backflow phase. 
 The shadings represent the statistical uncertainties of the results.
 Dotted lines: Integrated phase kinetic energy per unit length $\nu(r)=V_\Omega(r)/L_z$, where
 $V_\Omega(r)$ is the extra potential energy due to the vortex phase of the particles that lie 
 in the cylinder of radius $r$.
 Dot-dashed line: Fit of the MC data with the hydrodynamic form Eq.~\eqref{classical}. }
\end{figure*}
In Fig.~\ref{fig1} we show the integrated vortex energy per unit length 
$\varepsilon_v(r)=\left[ E_v(r)-E(r)\right]/L_z$ as a function of radial distance 
$r$ for the BF and OF phase.
$L_z$ is the box side along the vortex axis and $E_v(r)$ and $E(r)$ are, 
respectively, the energy of the particles that lie inside the cylinder of radius 
$r$ centered on the $z$-axis in the system with and without the vortex line.
Therefore $\varepsilon_v(r)$ represents the vortex excitation energy per unit 
length integrated up to the radial distance $r$.
We plot $\varepsilon_v(r)$ up to a distance of 8 \AA~ because beyond this distance
the vortex flow field is modified with respect to the hydrodynamic $1/r$ as 
discussed in the previous Section, so that $\varepsilon_v(r)$ for $r>8$ \AA~ does
not have physical meaning.
It is evident from Fig.~\ref{fig1} that at all densities the BF phase 
\eqref{BFphase} reduces sensibly the energy of the vortex line excitation with
respect to the OF phase; as expected, backflow mostly affects the integrated 
vortex energy within about 2 \AA. 
Beyond this distance the two phases are essentially the same and we found that the 
energy gap
$\Delta \varepsilon_v (r) = \varepsilon_v^{\rm BF}(r) - \varepsilon_v^{\rm OF}(r)$
for $r>2$\AA~ is almost constant within the statistical uncertainties.
The average of £$\Delta \varepsilon_v (r)$ over the range $2<r<3$ \AA~ is 
$\overline{\Delta \varepsilon_v}  = -0.62 \pm 0.02$ 
K at $\rho=0.02$ \AA$^{-3}$, $-0.59 \pm 0.01$ K at $\rho=0.0218$ \AA$^{-3}$,
$-0.93 \pm 0.01$ K at $\rho=0.026$ \AA$^{-3}$ and $-1.00 \pm 0.01$ K at $\rho=0.03$ 
\AA$^{-3}$.
If we take $\varepsilon_v(r)$ at $r=2$ \AA~ as a measure of the vortex core energy
it turns out that the core energy given by BF is less than half that of the OF 
phase.
More precisely, the ratio $\eta=\varepsilon_v^{\rm BF}(r)/\varepsilon_v^{\rm OF}(r)$
at $r=2$ \AA~ takes the values $\eta=0.35$, $0.52$, $0.37$ and $0.30$ at the four densities 
of our computations, going from the lowest to the highest.

Classically, the vortex energy in an incompressible fluid has a logarithmic 
dependence on $r$. 
For a vortex line with circulation $\kappa$ this classical energy per unit
length is usually written as 
\begin{equation}
 \label{classical0}
 \varepsilon_{\rm hyd}(r) = \frac{\kappa^2}{4\pi} m \rho \left[ \ln\frac{r}{a} + \delta \right]
\end{equation}
where $a$ is assumed as a core parameter and $\delta$ is related to the model
for the core.\cite{saff,donn}
For instance, one has $\delta=0$ for an hollow core model and $\delta=1/4$ for
a core of radius $a$ rotating at uniform angular velocity as a solid body (solid 
core model).\cite{donn,fet2}
The specific model chosen for the vortex core is a matter of taste unless one wishes
to obtain absolute data\cite{glab} and one goes outside the hydrodynamics of an 
incompressible fluid.
The parameters $a$ and $\delta$ are clearly not independent and Eq.~\eqref{classical0} 
can be written in the form
\begin{equation}
 \label{classical}
 \varepsilon_{\rm hyd}(r) = \frac{\hbar^2}{m}\pi\rho\ln\frac{r}{\lambda}\ .
\end{equation}
where the core parameter lambda is $\lambda=ae^{-\delta}$ and the quantum value 
$\kappa = h/m$ for the circulation has been used.
For a quantum vortex within the GP equation the energy can be written in the form 
\eqref{classical} only for $r$ large compared to the coherence length as discussed below.

\begin{table}
 \caption{\label{tab2} Parameter $\lambda$ from the fit of the quantum vortex energy per unit
          length with the hydrodynamic Eq.~\eqref{classical} in the region $4-8$ \AA, and values 
          of the parameter $\Lambda$, Eq.~\eqref{elle}, for the energy of a large vortex 
          ring obtained from the present calculations (SPIGS) compared 
          to the values obtained with SWF\cite{vit2} (SWF) and with the values obtained by 
          fitting the experimental data\cite{glab} (exp.).}
 \begin{ruledtabular}
  \begin{tabular}{cccccc}
   \multirow{2}{*}{$\rho$ (\AA$^{-3}$)} & \multirow{2}{*}{phase} &
   \multirow{2}{*}{$\lambda$ (\AA)} & \multicolumn{3}{c}{$\Lambda$ (\AA)} \\
                          &    &          &  SPIGS  &  SWF  & exp. \\
   \hline
   \multirow{2}{*}{0.020} & OF & 0.45(2) & 3.32(4) &       &      \\
                          & BF & 0.85(2) & 6.03(8) &       &      \\
   \multirow{2}{*}{0.0218}& OF & 0.25(1) & 1.87(2) & 2.72  & \multirow{2}{*}{5.98} \\
                          & BF & 0.67(2) & 5.1(1)  & 5.87  &      \\
   \multirow{2}{*}{0.026} & OF & 0.38(1) & 2.72(3) &       & \multirow{2}{*}{7.68} \\
                          & BF & 1.13(3) & 8.2(1)  &       &      \\
   \multirow{2}{*}{0.030} & OF & 0.46(2) & 3.27(4) &       &      \\
                          & BF & 1.05(2) & 7.62(5) &       &      \\
  \end{tabular}
 \end{ruledtabular}
\end{table}
Our results for the quantum vortex give $\varepsilon_v(r)$ that is a slowly 
increasing function of $r$ with some structure that can be connected to the oscillations of the
density profile as discussed below.
However, statistical uncertainties are too large to characterize in detail such structures.
Our results for the energy of the quantum vortex for $r$ not too small can be represented in a 
reasonable way by the functional form \eqref{classical} for a suitable choice of
$\lambda$.
A fit of $\varepsilon_v(r)$ with $\varepsilon_{\rm hyd}(r)$ over the range
$4-8$ \AA~ gives the values of $\lambda$ reported in Tab.~\ref{tab2} and the
corresponding $\varepsilon_{\rm hyd}(r)$ is plotted in Fig.~\ref{fig1} as a 
dot-dashed line.
For a vortex in bulk $^4$He at distances larger than $8$ \AA~ the contribution to the vortex 
energy due to interparticle correlations and to backflow should have decayed to zero and the 
density is uniform so that the only remaining contribution to $\varepsilon_v(r)$ is the kinetic 
energy due to the $1/r^2$ centrifugal barrier.
Therefore Eq.~\eqref{classical} is the correct representation of the vortex energy 
also at arbitrary large distance for a straight vortex in bulk $^4$He. 
It should be stressed again that the parameter $\lambda$ for the quantum vortex is 
just a convenient way to represent the energy of the quantum vortex such that it 
joins with the large scale behavior and $\lambda$ does not represent the core radius 
or the healing length.
In summary, both for the OF and the BF phase Eq.~\eqref{classical} is a good representation of 
the quantum $\varepsilon_v(r)$ starting from $r>4$ \AA, below this distance 
$\varepsilon_{\rm hyd}(r)$ does not represent the quantum energy and one should use the numerical 
results reported in Fig.~\ref{fig1}.

The vortex excitation energy $\varepsilon_v(r)$ can be decomposed into several contributions. 
One derives from the expectation value of the extra potential $V_\Omega(R)$ and this corresponds 
to the kinetic energy due to the phase of the wave function. 
Another contribution represents the extra kinetic energy due to the bending of the real part of 
the wave function close to the core. 
Finally there is a contribution due to the change of the expectation value of the Hamiltonian due 
to local rearrangement of the atoms close to the core. 
In Fig.~\ref{fig1} we plot also the contribution $\nu(r)$ of $V_\Omega$ to $\varepsilon_v(r)$. 
In the case of the OF phase,  $\varepsilon_v(r)$ is significantly larger of $\nu(r)$ so that a 
substantial contribution to the vortex core energy is due to the bending kinetic energy. 
It is interesting to notice that  $\varepsilon_v(r)$ is much closer to $\nu(r)$ in the case of 
the BF phase. 
Actually at freezing density and at $\rho=0.03$ \AA$^{-3}$, within the statistical uncertainty,
$\varepsilon_v(r)$ coincides with $\nu(r)$, i.e. BF is so efficient that it essentially cancels 
the bending energy. 
At the two lowest densities of our computations there is some remaining contribution from the 
bending energy. 
We conjecture that this might be due to the choice of the BF phase Eq.~\eqref{BFphase} that is more 
accurate at large density and less so at lower density.

\begin{table*}
 \caption{\label{tab1} Core radii $d_{\rm WHM}$ and $R_{\rm cyl}$ as defined in the text
  at different densities for both OF and BF phases.
  Integrated energy per unit length of a vortex line at 6 \AA~ from the core for the
  present computations (SPIGS) and for the variational approach of Ref.~\onlinecite{vit2}
  (SWF).
  Damping parameter $r_1$ in Eq.~\eqref{fitrho}.
  Wave vector $q_{\rm max}$ of the first peak of $\tilde\rho(q)$,
  roton wave vector $q_{\rm rot}$ from Ref.~\onlinecite{over} and position of the main peak, $q_{\chi}$,
  in the static density response function, $\chi(q)$,\cite{tomo} computed via the GIFT method.\cite{gift}
  $q_{\chi}$ has been obtained from a parabolic fit of $\chi(q)$ in the region of the main peak
  (data at $0.020$ and $0.030$ \AA$^{-3}$ are from unpublished computations).
  }
 \begin{ruledtabular}
  \begin{tabular}{cccccccccc}
   \multirow{2}{*}{$\rho$ (\AA$^{-3}$)} & \multirow{2}{*}{phase} & \multicolumn{2}{c}{core radius} &
   \multicolumn{2}{c}{$\varepsilon_v(r=6$~\AA$)$ (K/\AA)} & \multirow{2}{*}{$r_1$ (\AA)} &
   \multirow{2}{*}{$q_{\rm max}$ (\AA$^{-1}$)} &
   \multirow{2}{*}{$q_{\rm rot}$ (\AA$^{-1}$)} & \multirow{2}{*}{$q_{\chi}$ (\AA$^{-1}$)} \\
                          &    & $d_{\rm WHM}$ (\AA) & $R_{\rm cyl}$ (\AA) &   SPIGS   &   SWF  &     &      &      &                     \\
   \hline
   \multirow{2}{*}{0.020} & OF &  0.923  &     1.003     &  1.94(21) &    -   & 2.48 & 1.91(2) & \multirow{2}{*}{1.844(3)} & \multirow{2}{*}{1.936(2)} \\
                          & BF &  0.668  &     0.963     &  1.42(17) &    -   & 3.13 & 1.91(2) &                           &  \\
   \multirow{2}{*}{0.0218}& OF &  0.885  &     0.928     &  2.67(21) &  2.36  & 3.07 & 1.97(3) & \multirow{2}{*}{1.908(6)} & \multirow{2}{*}{1.989(3)} \\
                          & BF &  0.663  &     0.753     &  1.81(29) &  1.91  & 3.46 & 1.97(1) &                           &  \\
   \multirow{2}{*}{0.026} & OF &  0.821  &     0.606     &  2.74(27) &  2.97  & 3.40 & 2.09(1) & \multirow{2}{*}{2.048(6)} & \multirow{2}{*}{2.094(6)} \\
                          & BF &  0.565  &     0.456     &  1.51(33) &  1.91  & 3.89 & 2.18(1) &                           &  \\
   \multirow{2}{*}{0.030} & OF &  0.804  &     0.611     &  2.81(29) &   -    & 4.06 & 2.19(1) & \multirow{2}{*}{2.182(6)} & \multirow{2}{*}{2.192(4)} \\
                          & BF &  0.504  &     0.364     &  2.01(26) &   -    & 7.30 & 2.19(1) &                           &  \\
  \end{tabular}
 \end{ruledtabular}
\end{table*}

Classically the vortex energy \eqref{classical} is simply proportional to the
density if the vortex core parameter $\lambda$ is density independent.
We confirm the earlier finding\cite{sad1} of the SWF computation that the energy 
of a quantum vortex in superfluid $^4$He at short distance has very weak dependence 
on density.
As reported in Tab.~\ref{tab1} $\varepsilon_v(r)$ at $r=6$ \AA~ within the
statistical errors of our computation are the same at freezing and at equilibrium
density.

Within the mean field GP equation the quantum vortex energy deviates from the
hydrodynamic value \eqref{classical} at short distance because the local density 
$\rho(r)$ is a function of $r$.
From the numerical solution of the GP equation, accurate Pad\'e approximants for the
local density have been obtained.\cite{ber1,roar}
Using the approximant $[3/3]$ in Ref.~\onlinecite{roar} 
\begin{equation}
 \label{pade}
 f^2_{\rm GP}(r)=\frac{\rho(r)}{\rho}=\frac{0.3396r^2 + 0.0501r^4 + 0.0026r^6}{1+0.3976r^2+0.0527r^4+0.0026r^6}\ , 
\end{equation}
we have obtained the integrated vortex energy per unit length as
\begin{eqnarray}
 \label{GPvort}
 \varepsilon_v^{\rm GP}(r) &=& \pi\int_0^r \varrho d\varrho\left\{\left[f'_{\rm GP}(\varrho)\right]^2 
                               + \frac{f^2_{\rm GP}(\varrho)}{\varrho^2}\right. \nonumber \\ 
                           &&  \left. + \frac{1}{2}\left[1- f^2_{\rm GP}(\varrho)\right]^2\right\}\ ,
\end{eqnarray}    
where the standard GP reduced units are used.
In order to restore dimensional units for Eqs.~\eqref{pade} and \eqref{GPvort}, only 
a single parameter is required, i.e. the coherence (or healing) length $\xi$.
Unfortunately, for strongly correlated systems as liquid $^4$He, there is no 
unique definition for $\xi$.
The standard procedure is to choose $\xi$ such as the sound velocity provided by GP
equation is equal to the observed one.\cite{ber2}
This leads to $\xi=0.47$ \AA~ at the equilibrium density and $\xi=0.31$ \AA~ at the freezing one.
Authors of Ref.~\onlinecite{zucc} proposed another recipe by requiring the GP
core parameter (discussed below), which is about $1.5\xi$, to be equal to the 
experimental value,\cite{exp} resulting in $\xi=0.87$ \AA~ at the equilibrium density.\cite{not2}

\begin{figure}
 \includegraphics*[width=8cm]{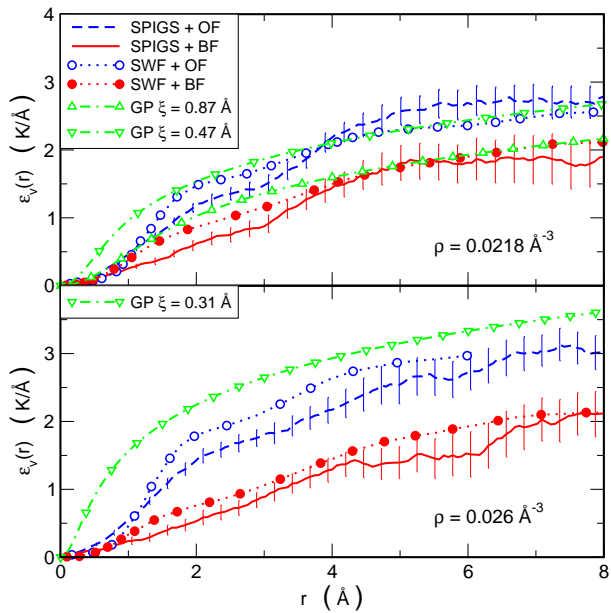}
 \caption{\label{fig2}
 (Color online) Comparison of integrated energy of a single vortex line per unit 
 length, $\varepsilon_v(r)$, at equilibrium (upper panel) and freezing (lower 
 panel) densities obtained with different methods.}
\end{figure}
In Fig.~\ref{fig2} the SPIGS vortex energy $\varepsilon_v(r)$ is compared to the GP results
for two choices for $\xi$ as well as with the SWF variational results.\cite{sad1}
There is an overall good agreement between the SWF results and the present fixed phase SPIGS
ones.
This is an additional evidence of the excellent quality of SWF in representing superfluid
$^4$He.
With respect to GP ones, one can say that it gives a reasonable representation of the vortex
energy, but its quality very much depends on the value of the ill-determined parameter $\xi$.
At equilibrium density the small value $0.47$ \AA~ gives $\varepsilon_v^{\rm GP}(r)$ in good
agreement with the exact SPIGS result for the OF phase.
What is surprising is that the GP result for $\xi= 0.87$ \AA~ is rather close to the SPIGS
result for BF phase.
At the freezing density, the agreement of QMC data with GP predictions is poor when $\xi$
is obtained in the standard way via the speed of sound.

Here a comment is in order.
With SWF the total energy $E_v$ of the vortex state and the energy $E_0$ of the ground state 
are each rigorously upper bounds to the exact values for the chosen model of the He-He interatomic
potential.
Though no such bound is present for the vortex ``excitation'' energy $E_v-E_0$.
On the other hand, in the fixed phase SPIGS computation both $E_v$ and $E_v-E_0$ are upper bounds
to the exact values because the SPIGS $E_0$ is exact within the statistical uncertainty.
Finally, no bound is present for the vortex energy given by the GP equation.

\begin{figure*}
 \includegraphics*[width=15cm]{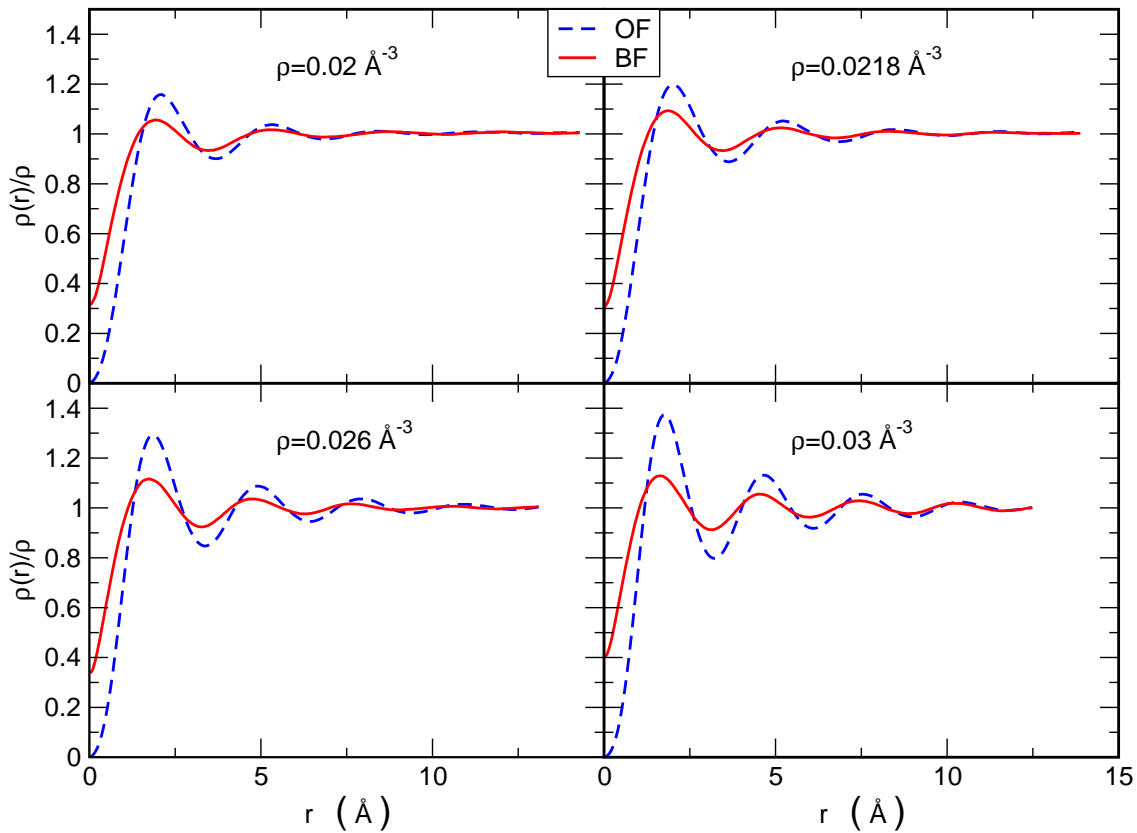}
 \caption{\label{fig3}
 (Color online) Radial density profiles, $\rho(r)/\rho$, for a single vortex line at 
 different densities in liquid phase.
 Dashed line: Onsager-Feynman phase. 
 Solid line: Backflow phase.}
\end{figure*}
In Fig.~\ref{fig3} we show the radial density profiles $\rho(r)$ obtained with the
OF and BF phases. 
In qualitative agreement with the pioneering work by Chester et al.\cite{ches}
$\rho(r)$ has a depression on the vortex axis and reaches the bulk density value 
in an oscillating way.
$\rho(r)$ for the OF phase vanishes on the vortex core.
At all studied densities $\rho(r)$ computed for the BF phase provides a partially
filled core in qualitative agreement with the SWF results. 
As it might be expected, the filling of the vortex core increases at large densities.
By comparing the BF and OF density oscillations around the vortex core one
can notice that such oscillations are significantly stronger for the OF phase and
shifted to larger distance compared to the BF. 
These oscillations can be well fitted with a damped oscillating function.\cite{dalf,vill}
Starting from $r\simeq 3$ \AA~ we have fitted our density profiles
$\rho(r)/\rho$ with the following functional form
\begin{equation}
 \label{fitrho}
 f(r) = f_0 + \frac{A}{\sqrt{r}}\cos(k_0r + \phi)e^{-r/r_1}
\end{equation}
with $f_0$, $A$, $k_0$ and $r_1$ fitting parameters.
The damping parameter $r_1$ is reported in Tab.~\ref{tab1}.
At all densities, the oscillations are more strongly damped when backflow is present.
It is clear that backflow has the role of reducing the perturbation effect of the
vortex and this explains the reduced core energy with respect to the OF phase.

As already pointed out, the core parameter is not rarely misled with the core radius,
but we stress again that the core parameter $\lambda$ is not a measure of the core extension, 
rather is a suitable value to be inserted in the hydrodynamic description Eq.~\eqref{classical} 
to obtain the large distance behavior of the vortex energy.
To obtain a measure of the vortex core extension is quite easy, even if not completely
unequivocal, within the GP approach since the density profile $\rho(r)$ is a smoothly 
increasing function of $r$.
The core radius is often taken equal to the radial distance where $\rho(r)/\rho=1/2$.\cite{zucc}
More difficult is to define it when $\rho(r)$ is an oscillating function of $r$ as 
the ones shown in Fig.~\ref{fig3}. 
As a possible choice we take $d_{\rm WHM}$ that we define as the position at which $\rho(r)$ 
is equal to the average of $\rho(r)$ at the first maximum and the value at the origin 
$r=0$. 
The values of $d_{\rm WHM}$ are given in Tab.~\ref{tab1}. 
Such a core radius for the BF phase is significantly smaller of that for the OF phase
and it shrinks for increasing density. 
We can obtain another way to measure the core radius in the following way. 
We might expect that at large distance from the core  $\rho(r)=\rho$, so that $f_0=1$ 
in Eq.~\eqref{fitrho}. 
We find that this is not so, the best fit of the computed $\rho(r)/\rho$ with 
Eq.~\eqref{fitrho} always gives $f_0$ slightly above 1, it is like as if the average 
density far from the vortex core were $f_0\rho$. 
We can understand this as a result of the combined effect of operating with a fixed 
number of particles, of finite size of the simulation box and of the expulsion of some 
particle from the vortex core.
In fact, in our simulations, the particles removed from the core accumulate in the 
region far from the vortex line, increasing the average density. 
We define $R_{\rm cyl}$ as the radius of an impenetrable hard cylinder, coaxial with 
the vortex axis and completely void of particles, that inserted in the same simulation 
box gives rise to a density increase equal to $f_0\rho$. 
Thus $R_{\rm cyl}$ is also a measure of how many particles are expelled from the core 
of the vortex.
The obtained values of $R_{\rm cyl}$ are given in Tab.~\ref{tab1}. 
$R_{\rm cyl}$ is of the same order of $d_{\rm WHM}$ and it has a similar dependence on 
the phase of the wf and on density.

\begin{figure}
 \includegraphics*[width=8cm]{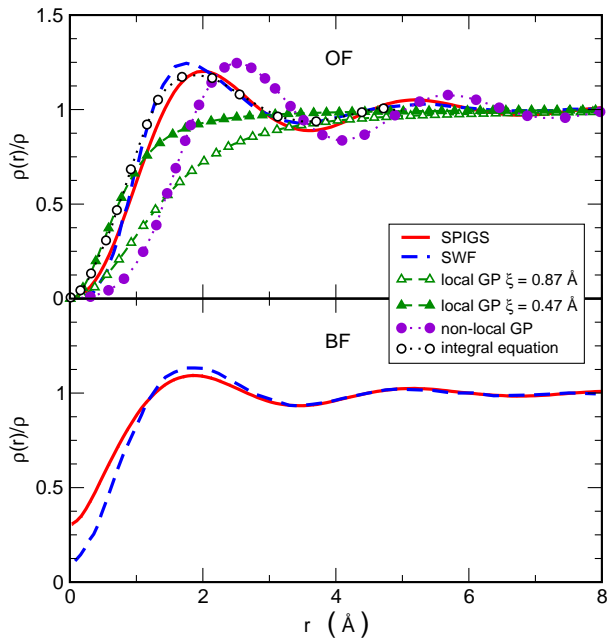}
 \caption{\label{fig4}
 (Color online) Comparison of normalized radial density profiles, $\rho(r)/\rho$,
 for a single straight vortex line at equilibrium density with Onsager-Feynman (upper
 panel) and backflow (lower panels) phases obtained with different methods.
 SPIGS results are from the present computation, SWF ones from Ref.~\onlinecite{sad1}.
 Non-local GP result is from the model in Eq.~(17) of Ref.~\onlinecite{ber2} (with
 $\gamma=1$, $\chi=3.5$ and $\delta=1$ in the notation of the original paper)
 and integral equation result is the one referred as B in Ref.~\onlinecite{ches}.}
\end{figure}
The fixed phase SPIGS density profile at the equilibrium density is compared in 
Fig.~\ref{fig4} to the results of other theories. 
In the upper panel the results for the OF phase are shown. 
One can notice that there is a good agreement of the SPIGS result with that of the 
variational SWF computation.\cite{sad1} 
As already noticed, the GP $\rho(r)$ is a monotonically increasing function of $r$, 
no oscillations are present. 
What is commonly called the GP equation is based, in addition to a mean field approximation, 
also on the assumption that the interatomic interaction $v(\vec r_1-\vec r_2)$ is a contact one, 
i.e. $v(\vec r_1-\vec r_2)$ is proportional to $\delta(\vec r_1-\vec r_2)$. 
If one relaxes  this assumption and takes a finite range $v(\vec r_1-\vec r_2)$ one gets a 
non-local GP equation (also called non-local non-linear Shr\"odinger equation) and 
$v(\vec r_1-\vec r_2)$ has the role of a phenomenological effective interatomic potential 
between He atoms.\cite{gros,rica} 
A further extension of GP has been studied in which a term proportional to a power higher 
than 4 of the single particle wf is included too.\cite{dupo} 
The density profile around a straight vortex line as obtained with one of the most
studied non-local GP models\cite{ber2} is also reported in Fig.~\ref{fig4}.
It indeed gives an oscillating $\rho(r)$ with oscillations of amplitude comparable to those 
of the SPIGS computation, but the positions of the extrema are quite off the mark (remember 
that we are comparing theories with the same OF phase and that the SPIGS result is exact). 
Also the integral equation approach of Ref.~\onlinecite{ches} gives an oscillating $\rho(r)$
which is an overall good agreement with QMC results.
In the lower panel of Fig.~\ref{fig4} the density profiles for two computations with backflow are 
shown, the present SPIGS and the SWF. 
The SPIGS result gives a larger population of the core and a somewhat smaller core size compared 
to the SWF result. 
Notice that the BF phase is not the same in these two computations, in the present computation 
the phase is explicit as given in Eq.\eqref{BFphase} and it is variationally optimized, whereas
in the SWF the phase is implicit because it derives from from a many-body integration over 
subsidiary variables.

\begin{figure*}
 \includegraphics*[width=15cm]{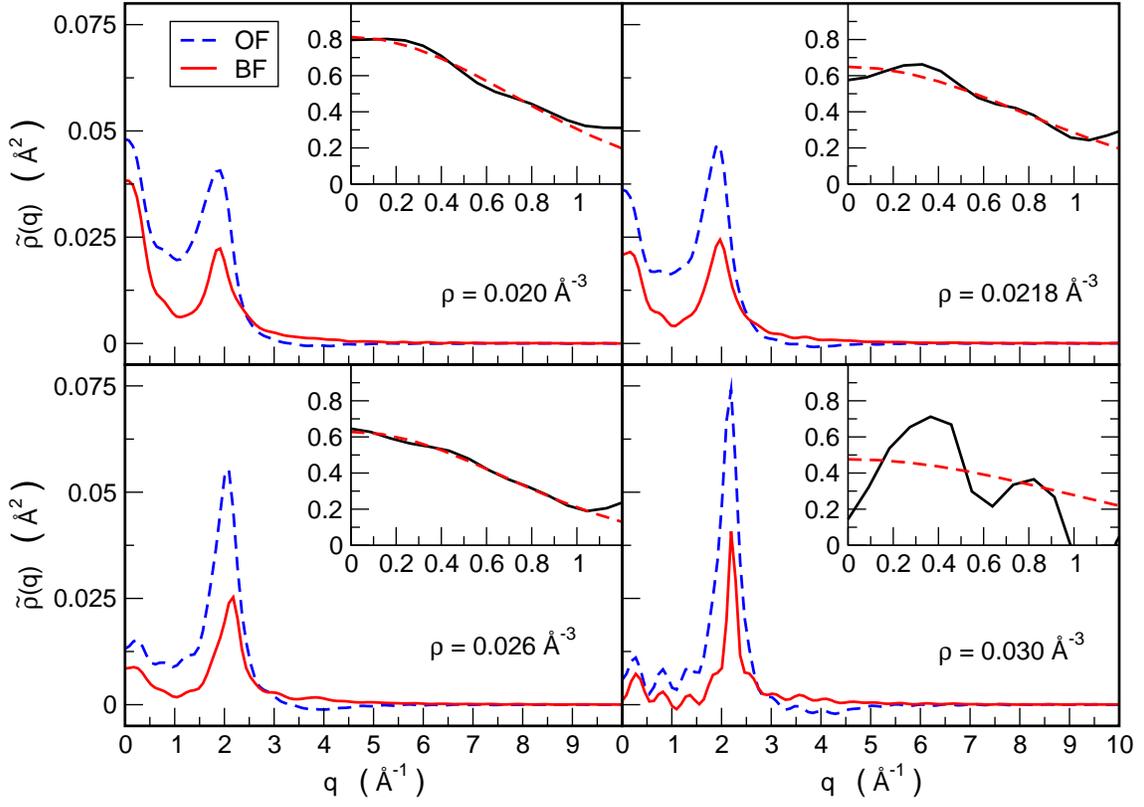}
 \caption{\label{fig5}
 (Color online) Fourier transform of the radial density profile, $\tilde\rho(q)$, for a
 single vortex line at different densities in liquid phase. 
 Dashed line: Onsager-Feynman phase. 
 Solid line: Backflow phase.
 Insets: ratio between the Fourier transforms of the radial density profile with OF
 phase and with BF phase
 (solid line) and Gaussian fit in the $0<q<1$ \AA$^{-1}$ range (dashed line).} 
\end{figure*}

In the contest of density functional theory it has been proposed\cite{vill,dalf}
that the oscillations of the local density around the vortex core could be
related to the roton excitation. 
The range of densities considered in the present work corresponds to the range
of densities recently studied\cite{over} in the characterization of the 
excitation spectrum via the exact SPIGS method and GIFT,\cite{gift} a novel 
powerful approach to extract dynamic structure factors from imaginary time density 
correlation functions.
We have, therefore, the possibility to investigate such a relation on the
basis of a fully microscopic approach.
In order to estimate the wave vector associated to the density oscillation around
the vortex core we have computed the Fourier transform, $\tilde\rho(q)$, of the radial
density profile $\rho(r)/\rho - 1$.
In computing $\tilde\rho(q)$, the SPIGS data have been extended with the function 
$f(r) - f_0$, from \eqref{fitrho}, beyond $L/2$.
The obtained $\tilde\rho(q)$ are shown in Fig.~\ref{fig5}.
At each studied density a well defined peak is present in $\tilde\rho(q)$; its position,
$q_{\rm max}$, indicates the wave vector which characterizes the oscillations in
the radial density profile.
The obtained $q_{\rm max}$ are reported in Tab.~\ref{tab1} together with the
roton wave vectors,\cite{over} $q_{\rm rot}$, and the position of the main peak, 
$q_\chi$, of the static density response function,\cite{tomo} $\chi(q)$, at the 
considered densities.
We find that $q_{\rm max}$ is essentially the same for the OF and the BF phases,
suggesting that it might be related to an intrinsic property of the bulk system.
We find that even if they are very similar, $q_{\rm max}$ is always higher than 
the roton wave vector $q_{\rm rot}$, rather, we find a much better agreement with 
$q_\chi$.
This is not surprising since 
$\chi(q)$ provides the amplitudes of the 
density response of the system to a static perturbation, which is the case of a
vortex within fixed phase approximation. Then $q_\chi$ corresponds
to the preferred wave vector for a density modulation.

A vortex line has excited states in the form of Kelvin waves in which the vortex 
is no more straight but its core moves in a helical way.
One way of interpreting the delocalization of vorticity achieved with BF phase is
that it is due to the zero point motion of such Kelvin waves.
In fact, the localized vorticity given by the OF phase should spread out in a 
cylindrical region around the vortex axis as a consequence of such zero-point 
motion.
Under such hypotesis the density profile $\rho_{\rm BF}(r)$ of the BF
phase should be equal to the convolution of the density profile $\rho_{\rm OF}(r)$
of the OF phase and of the probability $P(r)$ that the vortex core has a 
transverse displacement equal to $r$.
For harmonic oscillations $P(r)$ should be a Gaussian function.
One can easily verify if $\rho_{\rm BF}(r)$ indeed can be represented in this way. 
In fact, in Fourier space the convolution becomes a product, so the ratio of the 
Fourier transforms of $\rho_{\rm BF}(r)$ and $\rho_{\rm OF}(r)$ should be a gaussian 
functions of the wave vector $q$.
This ratio is plotted in the insets in Fig.~\ref{fig5} as well as the best fit by
a Gaussian function over the range $0-1$ \AA$^{-1}$.
It is clear that indeed the ratio $\tilde\rho_{\rm BF}(q)/\tilde\rho_{\rm OF}(q)$ is to a 
good approximation a Gaussian and this give support to the notion that backflow 
for a vortex is a way to represent the zero point motion of Kelvin waves.
The optimal fitting values for the variance of the Gaussian are $0.712\pm0.016$
\AA~at $0.020$ \AA$^{-3}$, $0.777\pm0.044$ \AA~at $0.0218$ \AA$^{-3}$, 
$0.676\pm0.010$ \AA~at $0.026$ \AA$^{-3}$ and $0.963\pm0.016$ \AA~at $0.030$ 
\AA$^{-3}$.
A significant deviation of $\tilde\rho_{\rm BF}(q)/\tilde\rho_{\rm OF}(q)$ from a gaussian is
present only at the highest density $\rho = 0.030$ \AA$^{-3}$ in the metastable
fluid phase.
We do not know if this is a genuine effect or if it is a consequence of size 
effects that are more pronounced at higher density due to the slow decay of the 
density oscillations.

\section{Large vortex ring in superfluid $^4$He}
\label{s:ring}

Vortex ring excitations are particularly important for superfluid $^4$He since
experimental data\cite{exp} on mobility of ions trapped in the core of a vortex 
ring have given information on such excitations.\cite{exp,glab}
A full description of a vortex ring requires a more complex functional form for 
the phase, but for large vortex ring, when the radius $R$ is much larger than 
the core size, the energy is expected to be accurately approximated by the sum of 
the kinetic energy of an incompressible flow outside a toroidal region centered 
at the circle of radius $R$ and with a radius $b$, and of the energy of the core 
inside this region.\cite{ches}
This latter energy can be approximated by $2\pi R\varepsilon_v(b)$, 
where $\varepsilon_v(b)$ is the energy per unit length of a straight vortex line 
up to a distance $b$.\cite{ches,vit2}
Here $b$ represents the distance at which the inner quantum flow field is joined 
to the external hydrodynamic one and should not be confused with the core size or the 
healing length.\cite{vit2}
For the outer region the energy is the one of a classical vortex ring of 
radius $R$ and a hollow core of size $b$:
\begin{equation}
 E(R,b) = \frac{1}{2} \kappa^2 m \rho R \left[ \ln\frac{8R}{b} - 2 \right] 
\end{equation}
where the circulation has been set to its quantum value $\kappa=h/m$.
Then, within such an approximation, the excitation energy of a large vortex ring
reads
\begin{equation}
 \label{vortexring}
 E_{\rm ring}(R) = \frac{h^2}{2m}\rho R\left[\ln\frac{8R}{b} - 2 + 
                   \frac{4\pi m}{h^2\rho}\varepsilon_v(b) \right].
\end{equation}
 
The experimental results of Rayfield and Reif\cite{exp} have been interpreted\cite{glab}
in terms of an energy of a hollow-core vortex ring of radius $R$, written in the form
\begin{equation}
 \label{vortexring2}
  E_{\rm ring}(R) = \frac{h^2}{2m}\rho R \ln\frac{8R}{\Lambda}.
\end{equation}
Our approximated theoretical expression \eqref{vortexring} can be recast in the
form of \eqref{vortexring2} with the core parameter
\begin{equation}
 \label{elle}
 \Lambda = be^{2-\frac{4\pi m}{h^2\rho}\varepsilon_v(b)}\ .
\end{equation}
Our values of the parameter $\Lambda$ are reported in Tab.~\ref{tab2} together with the 
values obtained by the fit over the experimental data.\cite{glab}
In order to account for statistical uncertainty we have averaged Eq.~\eqref{elle}  
over the range $4<b<8$ \AA~ where the computed energy per unit length is quite close
to the hydrodynamic one \eqref{classical}.
If the dependence of $\varepsilon_v(b)$ on $b$ is well represented by
Eq.~\eqref{classical}, the energy of the ring \eqref{vortexring} turns out to be
independent on the choice of $b$,\cite{vit2} and the parameter $\Lambda$ is equal to 
$e^2\lambda$.
Our results excellently fulfill such a relation, with deviation smaller than 4\%,
confirming that, for large radial distances, the energy per unit length of a 
vortex line is given with great accuracy by the classical hydrodynamic
functional form \eqref{classical}.

We stress again that the parameters $\lambda$ in Eq.~\eqref{classical} and $\Lambda$ 
in \eqref{vortexring2} do not represent the GP coherence length or the vortex core radius.
Failure to recognize this can lead to an incorrect interpretation of experimental data.
For instance, the interpretation of the experimental data in terms of classical 
hydrodynamics\cite{glab} led to a core parameter $0.81$ \AA~ and $1.04$ \AA~ respectively 
at equilibrium and at freezing density.  
This expansion of the core parameter with the density has been addressed as the test
bed for any successful theory of the core structure, and a vortex model that fulfills 
this requirement has actually been developed.\cite{glab}
However, if we consider the core parameter as a measure of the core extension we will be 
in the counterintuitive situation of a vortex that expands by increasing the density.
This striking feature is due to a misleading interpretation of the hydrodynamic 
model \eqref{classical}.
In fact, as shown in Tab.~\ref{tab2}, our theory with the BF phase gives at both 
equilibrium and freezing density results in rather good agreement with the experimental 
value of $\Lambda$, i.e. an expanding core parameter; but the actual core size, as measured 
by the density profile (parameters $d_{\rm WHM}$ and $R_{\rm cyl}$ in Tab.~\ref{tab1}), 
decreases by about 15\% on going from equilibrium to freezing density.
Moreover the vortex rings investigated in the experiments\cite{exp,glab} have a radius 
in the decade $10^3-10^4$ \AA~ and it might be assumed that for such large vortices a
full classical description should be adequate. 
In order to see that this is not correct consider the terms in square parentheses in 
Eq.~\eqref{vortexring}. 
The factor $\frac{h^2\rho}{4\pi m}$ is $0.83$ K/\AA~ at equilibrium density and using 
$\varepsilon_v(b)$ for $b=6$ \AA~ from Tab.~\ref{tab1} we find that the third term 
in square parentheses in \eqref{vortexring} is equal to $2.12$. 
This term is equal to $1.51$ when we perform a similar computation at freezing density. 
Thus if one does not take into account the variation with density of the quantum energy 
one gets an incorrect description of the vortex.

\section{A model for a vortex line in solid $^4$He}
\label{s:solid}

Given the recent interest on supersolidity and on the possible presence of 
vortex-like excitations in solid $^4$He,\cite{ande,kubo1,kubo2,kim1,kim2}
we have extended our study also to the solid phase. 
Exact quantum Monte Carlo computation at finite\cite{prok,cipo}
and zero temperature\cite{miarev} have shown that an ideal perfect
$^4$He crystal (the so called commensurate solid) has no BEC and is not superfluid.
Without the presence of BEC the standard order parameter cannot be defined and the
form \eqref{OFphase} or \eqref{BFphase} for the phase of a quantum vortex has little
justification because such a state should rapidly decay into other excitations.
On the other hand it is known that almost any deviation from the ideal perfect state
does induce BEC in solid $^4$He.\cite{vaca,grai,disl,fate}
The idea was to check the possibility that an ideal perfect $^4$He crystal could 
correspond to a marginally stable quantum system,\cite{miarev} in which rotation 
would be a strong enough perturbation to induce a ``dynamical'' 
Bose--Einstein condensation and quantum vortices at the same time.
In this case the OF and BF variational ansatz on the phase of the many--body wf 
could be justified for an order parameter defined at least locally.\cite{shira}
Since the BF corrections have a larger effect on the core filling as the density is
increased in the liquid phase, one could expect that BF should be even more relevant
in the solid phase.

\begin{figure}
 \includegraphics*[width=9cm]{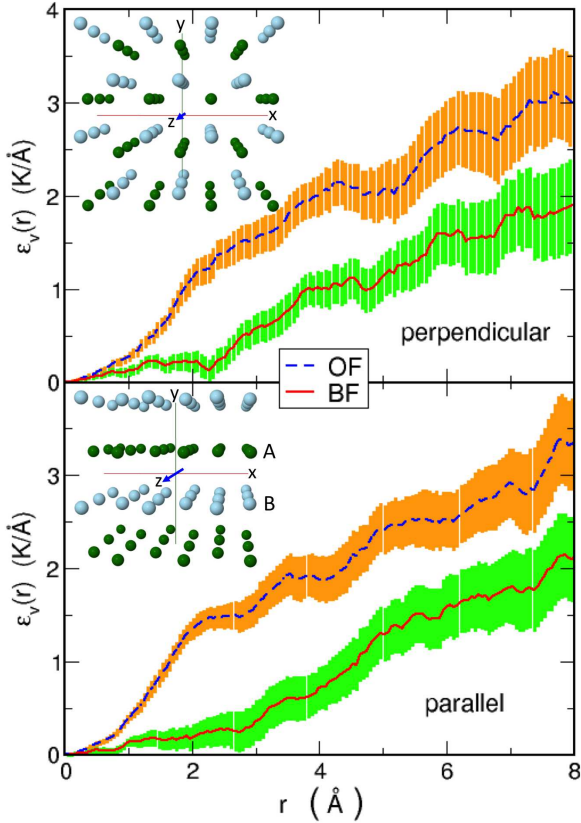}
 \caption{\label{fig7}
 (Color online) Integrated energy of a vortex line per unit length, 
 $\varepsilon_v(r)$, at $\rho=0.0293$ \AA$^{-3}$ in a hcp solid as 
 function of the distance from the vortex core, for two different orientation of 
 the vortex axis (which coincides with the $z$ axis): perpendicular to the basal 
 planes (upper panel) and parallel to the basal planes (lower panel) as shown in 
 the insets where the vortex axis points toward the reader along the $z$ axis.
 Atoms belonging to different basal planes in the hcp structure (A and B) have 
 been represented with different colors.
 Dotted line: Onsager-Feynman phase. 
 Solid line: Backflow phase. 
 The shadings represent the statistical uncertainties of the results.}
\end{figure}
\begin{figure}
 \includegraphics*[width=8cm]{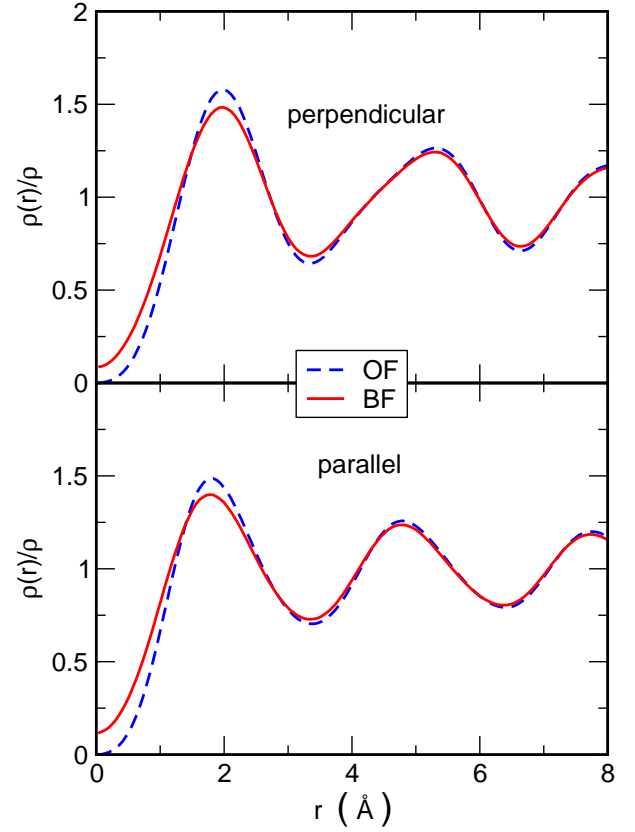}
 \caption{\label{fig8}
 (Color online) Normalized radial density profiles,$\rho(r)/\rho$,
 at $\rho=0.0293$ \AA$^{-3}$ in a hcp solid as functions of the distance from the 
 vortex core, for two different orientation of the vortex axis: perpendicular to the 
 basal planes (upper panel) and parallel to the basal planes (lower panel).
 Dotted line: Onsager-Feynman phase.
 Solid line: Backflow phase.}
\end{figure}
We have thus performed fixed phase SPIGS computation for the OF \eqref{OFphase} and the 
BF \eqref{BFphase} phase also in bulk solid $^4$He.
In Fig.~\ref{fig7} we report the integrated energy per unit length 
$\varepsilon_v(r)$ and in Fig.~\ref{fig8} the radial density profile 
$\rho(r)/\rho$ as a function of the core distance $r$ for the solution of 
Eq.~\eqref{schremod} in an ideal perfect hcp $^4$He crystal at $\rho=0.0293$
 \AA$^{-3}$, just above the melting density.
We have considered two possible orientations for the vortex axis: perpendicular to 
the basal planes (i.e. along the $c$-axis) and parallel to the basal planes.
As in the liquid phase, the BF phase provide a energy gain of about 1 K with 
respect to the OF one, but the most striking result is that the core is partially 
filled even in the solid phase when the BF phase is considered. 
The filling of the core is however smaller than what found in the liquid.
Within the statistical uncertainty the obtained $\varepsilon_v(r)$ are the same 
for the two considered orientations of the vortex axis both with the OF and with
the BF phase.

An interesting information that we can derive from our computation is the 
deformation of the crystalline lattice and the location of the vortex axis with
respect to the crystalline lattice sites.
In fact, in a SPIGS computation the equilibrium positions of the atoms are generated
as a spontaneous broken symmetry and the crystal is free to translate.
We find that in the simulation, whatever is the starting configuration of the atoms
with respect to the vortex axis the system evolves into a state in which the vortex
axis is an interstitial line, in the sense that this straight line lies as far as 
possible from the lattice sites as shown in the insets in Fig.~\ref{fig7}.
A similar result was already noticed in two dimensional $^4$He crystals.\cite{2Dvor}
In addition we find that the centrifugal barrier induces only a very small distortion
of the crystal lattice, again in agreement with the result in 2D \cite{2Dvor}.
The vortex core filling is greater in the case where the axis of the centrifugal flow
field is parallel to the basal planes; we have found, in fact, that in this case the 
solid is free to slide with respect to the axis and this gives rise also to a less 
structured density profile around core axis.

\begin{figure}
 \includegraphics*[width=8cm]{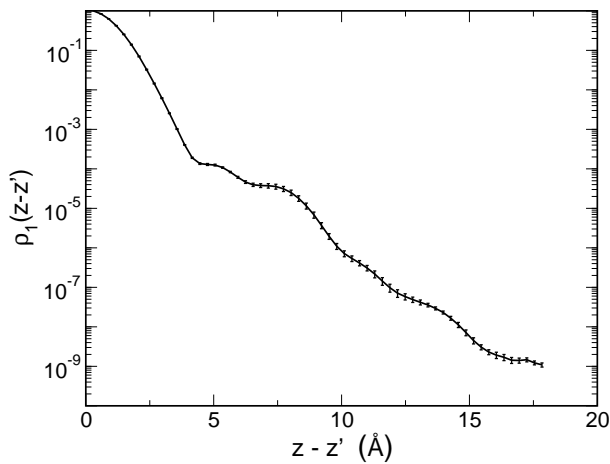}
 \caption{\label{figobdm}
 One--body density matrix computed in a perfect hcp $^4$He crystal at $\rho = 0.0293$ \AA$^{-3}$
 along the direction parallel to the axis of the centrifugal flow field of the OF phase.
 The case shown here corresponds to an axis perpendicular to the basal planes of the hcp crystal.
 }
\end{figure}
We have also computed the one--body density matrix $\rho_1(\vec r,\vec r')$ in the
system along the axis of the centrifugal flow field.
We find that for both phases $\rho_1(\vec r,\vec r')$ decays exponentially with increasing 
$|\vec r - \vec r'|$ as in the ideal perfect crystal.
As an example, in Fig.~\ref{figobdm} we show the result obtained for the one--body density matrix 
in presence of a centrifugal flow field related to the OF phase with the axis perpendicular
to the basal planes of a hcp $^4$He crystal.
A very similar result is obtained with the BF phase.
We conclude that both the OF and the BF variational ansatz are not able to induce any 
off--diagonal long--range order in the crystal.
This means that there is no condensate fraction induced by rotation and that the state
corresponding to the solution of Eq.~\ref{schremod} is not a stable quantum state
in an ideal perfect $^4$He crystal.
It should be noticed that it has been argued\cite{and2} that in solid $^4$He 
vortices can appear also in presence of a strongly fluctuating order parameter 
such that no phase coherence is present.
Also in this case, the assumed forms \eqref{OFphase} and \eqref{BFphase}
of the phase has little justification, so we cannot investigate this possibility with 
the present computation.

\section{Conclusions} 
\label{s:concl}

We have performed a microscopic characterization of a single straight vortex line in three
dimensional $^4$He systems. 
By using unbiased quantum Monte Carlo methods at zero temperature and the fixed--phase 
approximation, we have obtained an exact estimation of the local energy and the local density 
profile around the vortex core once the Feynman--Onsager model or a phase with backflow 
correlations is assumed.
The present results give a much stronger theoretical basis to some of the earlier findings 
obtained by the variational SWF method,\cite{vit2,sad2} like the large lowering of the 
vortex core energy due to BF correlations, the weak dependence of the vortex core energy on 
density or the significant oscillations of the density profile close to the core. 
In fact our results represent rigorous upper bounds of the exact excitation energy of a vortex. 
For the OF phase it is not possible to do better of our result because we solve exactly the 
quantum problem for the real part of the wf. 
With backflow our results depend on the specific choice of the BF phase so it could be improved. 
In any case our results allow us to conclude that it is not possible to get a correct quantitative 
estimation of the vortex core energy without taking into account backflow correlations. 
In fact the vortex excitation energy integrated up to a radial distance of order $2-4$ \AA~ is 
reduced by about a factor of 2 with respect to the value for the OF phase so any better BF phase 
would reduce that energy even more. 
The present results for a straight vortex line has an important implication for vortex rings of 
radius much larger of the core radius and on interpretation of experimental data.  
It is customary to represent the energy and the translational velocity of the ring in term of 
formulae derived from hydrodynamics with the core parameter determined by the experimental 
data. 
We show that the energy of a vortex ring can indeed be represented with the hydrodynamic formula 
but the length parameter contained in it is not a measure of the size of the core
but, as already shown in Ref.\onlinecite{sad1}, it contains information on the core energy and this       
requires the quantum theory.
Our results for the energy of the ring as function of its radius are in agreement with experiments 
both at equilibrium density and at freezing, at the same time the core radius shrinks for 
increasing density and the core parameter $\Lambda$ expands. 
We remark that an increasing value of $\Lambda$ with pressure, 
as obtained when the experimental data are interpreted by using the hydrodynamic description,
is not a signature of an expanded vortex core. 

Backflow not only reduces the excitation energy but also gives smaller oscillation of the density 
profile compared to the OF case and the density is finite even on the core axis. 
Analysis of the density profile with and without backflow shows that the first can be represented 
by a convolution of the second with a gaussian function. 
This is suggestive that BF is a way to represent the effect of the zero-point motion of Kelvin waves 
and from the width of the gaussian we estimate a mean square oscillation of about $0.7$ \AA.  
This suggestion is worthy of further study. 
In Fourier space the density profile has a strong peak at a wave vector that at all densities is 
very close to the wave vector at which the static density response function has a maximum. 
This wave vector is somewhat larger of the one at the minimum of the roton dispersion curve so that 
our microscopic computations partially modify previous suggestions\cite{dalf,vill} based on 
phenomenological theories. 
The fact that the Fourier spectrum  of the density profile is sharply peaked in $q$ in the region of 
rotons with positive group velocity (so called R$^+$ rotons) gives a hint that in a vortex 
reconnection event there might be emission of rotons\cite{ogaw} in preference of phonons as obtained
on the basis of GP equation.\cite{zucc} 
Study of vortex reconnections with QMC methods to verify this prediction seems presently out of reach 
of microscopic simulation. 
It might be feasible to extend the GP study of reconnection to one of the non local GP 
equations. 
For instance the model by Berloff and Roberts \cite{ber2}  is attractive because it gives a 
realistic description of the energetics of the roton excitations as well as a reasonable density 
profile around the vortex core as shown in Fig.~\ref{fig4}. 
Should roton emission in a vortex reconnection depend more on the modulation of the density
profile of the vortex than on the delocalized vorticity, it will be interesting to study the 
vortex reconnection with such an equation because it should give evidence for roton generation 
in a vortex reconnection.
It has been also suggested in the literature\cite{gla2} that around the vortex axis there
might be a condensation of rotons.
We do not find a way to prove or disprove such hypothesis on the basis of the present microscopic
theory.

We have studied the vortex excitation also in metastable states below equilibrium density and above 
freezing. 
We do not find evidence of any anomaly for the studied densities, the vortex properties have a smooth 
density dependence within the statistics of our simulations.

A number of other properties of a vortex can be calculated with our simulation method.
One is the computation of the flow field and of vorticity in the core region. 
This will allow to compute the translational velocity of a large vortex ring with BF phase in a way 
similar to what was done in Ref.~\onlinecite{ches} for the OF model.
Another interesting question is what happens to the Bose-Einstein condensate fraction $n_0$ in the core. 
It has been suggested\cite{gla2} that in the core the fluid is normal even at $T = 0$ K.             
The only microscopic computation of the condensate for a vortex in superfluid $^4$He is based on the 
integral equation method\cite{ches} and the result was an increasing condensate fraction $n_0/\rho$ 
in the core but both $n_0$ and the local density were vanishing at the vortex axis because that 
computation was based on the OF phase. 
With BF phase the core is partially filled and it will be very interesting to compute $n_0$ with the 
present method. 
We leave such computations for future work.
A further challenging development is the direct microscopic simulation of a small vortex ring.

We have investigated in crystalline solid $^4$He a state with the same forms of the vortex phase as in 
the liquid phase.
We find that the vortex core goes into interstitial positions and that there is a rather weak lattice 
distortion. 
At the same time we find that the system has no off-diagonal long-range order like in the perfect solid, 
i.e. the perturbation introduced by the phase is not able to induce a condensate. 
Therefore the studied excited state should not be a good representation of an elementary excitation of 
the solid. 
The topic of a microscopic model of the proposed vortex excitation in crystalline $^4$He in presence of 
a strongly fluctuating local order parameter\cite{ande} remains to be explored.

\acknowledgments
We acknowledge the CINECA and the Regione Lombardia award, under the LISA initiative,
for the availability of high performance computing resources and support.
One of us (LR) wants to thank Dipartimento di Fisica, Universit\`a degli Studi 
di Milano, for some support to his research activity.


\begin{thebibliography}{99}
\bibitem{note} Authors of Ref.~\onlinecite{gior} found an hollow core even with
               BF correction because their trial wave function maintains a 
               divergence of the velocity field at the vortex core.
\bibitem{not2} Authors of Ref.~\onlinecite{zucc} used as experimental value for the core
               parameter $a$ the one given in Ref.~\onlinecite{exp}.
               This value has been obtained from the experimental results for energy and
               velocity of vortex rings via a solid core model keeping the quantum of
               circulation $\kappa$ as a fitting parameter.
               Authors of Ref.~\onlinecite{glab} provide a different set of values by using
               an hollow core model and with $\kappa$ set to its theoretical value.
               We shall compare our results with this second set of data, since even in our 
               computation $\kappa$ is fixed to its theoretical value.
\end{thebibliography}
\end{document}